\input psfig 
\documentstyle[aas2pp4,tighten]{article}
\newcommand{\beq}{\begin{equation}}
\newcommand{\eeq}{\end{equation}}

\newcommand{\kms}{km s$^{-1}$~}

\begin{document}

\title{Precise Masses for Wolf 1062 AB \\from Hubble Space Telescope
 Interferometric Astrometry \\and McDonald Observatory Radial Velocities \footnote{Based on 
observations made with
the NASA/ESA Hubble Space Telescope, obtained at the Space Telescope
Science Institute, which is operated by the
Association of Universities for Research in Astronomy, Inc., under NASA
contract NAS5-26555} }

\author{ G.\ F. Benedict\altaffilmark{1}, B. E.
McArthur\altaffilmark{1},
O. G. Franz\altaffilmark{2}, L.\ 
H. Wasserman\altaffilmark{2}, T. J. Henry\altaffilmark{3}, T. Takato\altaffilmark{6}, I. V. Strateva\altaffilmark{13}, J. L. Crawford\altaffilmark{14}, P. A. Ianna\altaffilmark{12}, D. W. McCarthy\altaffilmark{4} E.\ Nelan\altaffilmark{5}, W.\ H.\ Jefferys\altaffilmark{6}, W.~van~Altena\altaffilmark{7}, P.~J.~Shelus\altaffilmark{1},
 P.D. Hemenway\altaffilmark{8}, R. L.
Duncombe\altaffilmark{9}, D. Story\altaffilmark{10}, A.\ L.\
Whipple\altaffilmark{10}, A. J. Bradley\altaffilmark{11}, and L.W.\
Fredrick\altaffilmark{12}}

\altaffiltext{1}{McDonald Observatory, University of Texas, Austin, TX 78712}
\altaffiltext{2}{Lowell Observatory, 1400 West Mars Hill Rd., Flagstaff, AZ 86001}
\altaffiltext{3}{Dept. of Physics and Astronomy, Georgia State University, Atlanta, GA 30303-3083}
\altaffiltext{4}{Steward Observatory, University of Arizona}
\altaffiltext{5}{Space Telescope Science Institute, 3700 San Martin Dr., Baltimore, MD 21218}
\altaffiltext{6}{Astronomy Dept., University of Texas, Austin, TX 78712}
\altaffiltext{7}{Astronomy Dept., Yale University, PO Box 208101, New Haven, CT 06520}
 \altaffiltext{8}{Oceanography, University of Rhode Island, Kingston, RI 02881}
\altaffiltext{9}{Aerospace Engineering, University of Texas, Austin, TX 78712}
\altaffiltext{10}{ Jackson and Tull, Aerospace Engineering Division
7375 Executive Place, Suite 200, Seabrook, Md.  20706}
\altaffiltext{11}{Spacecraft System Engineering Services, PO Box 91, Annapolis Junction, MD 20706}
\altaffiltext{12}{Astronomy Dept., University of Virginia, PO Box 3818, Charlottesville, VA 22903}
\altaffiltext{13}{Dept. Astrophysical Science, Princeton University, Princeton, NJ 08544-1001}
\altaffiltext{14}{Applied Research Laboratories, University of Texas,
   Austin,   TX  78713-8029}


\begin{abstract}
We present an analysis of astrometric data from
FGS 3, a white-light interferometer on {\it HST}, and of radial velocity
data from two ground-based campaigns. We model the astrometric and radial velocity measurements simultaneously to obtain parallax, proper motion and component masses for
Wolf 1062 = Gl 748 AB (M3.5V). To derive the mass fraction, we relate FGS 3 fringe
scanning observations of the science target to a reference
frame provided by fringe tracking observations of a
surrounding star field. 

We obtain an absolute parallax $\pi_{abs} = 98.0 \pm 0.4$ milliseconds of arc, yielding ${\cal M}_A = 0.379 \pm 0.005{\cal M}_{\sun}$
and ${\cal M}_B= 0.192 \pm 0.003 {\cal M}_{\sun}$, high quality component masses with errors of only 1.5\%.

\end{abstract}


\keywords{astrometry --- interferometry --- stars: individual (Wolf 1062) --- stars: binary --- stars: radial velocities --- stars: late-type --- stars: distances --- stars: masses}


%

\section{Introduction}

This is a paper in a series presenting accurate  masses of stars on the lower main sequence, where the complex interplay of luminosity, age, and mass can be explored.
The dependence of intrinsic brightness upon mass, the
mass-luminosity relation (MLR), is applicable to many
areas of astronomy. After the H-R diagram,
it is perhaps the next most important relationship in stellar
astronomy, because the entire evolution of a star depends on mass.
The MLR remains poorly
determined for M dwarfs, by far the dominant population of the
Galaxy in both numbers ($>$70\%) and stellar mass contribution 
($>$40\%; Henry 1998), and is critical in delineating stars from brown dwarfs. Below 0.1 ${\cal M}_{\sun}$ an accurate mass determination can convincingly turn a brown dwarf candidate into a bonafide brown dwarf.

Wolf 1062 is an M dwarf binary with one component suspected to be in the 20-20-20 club, defined by Henry et al (1999) as having d $\le$ 20 pc, ($\pi \ge  $ 0\farcs05), ${\cal M} \le 0.2{\cal M}_{\sun}$, and orbital period P $\le$ 20 years. With the motivation  that the secondary is one of
the few objects that promises an accurate dynamical mass less than $0.2{\cal M}_{\sun}$, we obtained astrometric observations at 15 epochs with 
{\it HST} and radial velocities at 15 epochs with which to obtain a precise parallax and masses. Our astrometric observations  were
obtained with Fine Guidance Sensor 3 (FGS 3), a two-axis, white-light
interferometer  aboard {\it HST}. Most of our radial velocities were obtained at McDonald Observatory. Three velocities of component A came from \cite{Mar89}.

Table~\ref{tbl-1} provides aliases and physical parameters for Wolf 1062. 
We time-tag our data with a modified Julian Date, $mJD = JD - 2444000.5$, and abbreviate millisecond of arc, mas, throughout.

\cite{Bra91}
provide an overview of the
FGS 3 instrument and \cite{Ben99} describe the fringe tracking (POS) mode astrometric capabilities 
of FGS 3 and typical data acquisition and reduction strategies. Our response to the added astrometric complication of resolved orbital motion is discussed in Benedict et al. (2000b). Details concerning the treatment of fringe scanning (TRANS) mode data are available in a paper presenting a preliminary relative orbit for Wolf 1062 (\cite{Fra98}). To summarize, TRANS data are used to obtain
relative positions of the A and B components; POS data are used to obtain the
photocenter position of component A relative to a reference frame.

\cite{Fra99} present a definitive relative orbit obtained from fringe scanning (TRANS) data. The present analysis utilizes only the 15 (of 17 total) TRANS measurements where the secondary was resolved on both FGS axes. We combine these with 15 epochs of fringe tracking (POS) measurements relative to a reference frame and radial velocities to obtain a parallax, absolute orbits, and masses. POS data were not acquired at the first two epochs of TRANS data.
In contrast to our recent work on Gl 791.2 (Benedict et al. 2000b), we include ground-based radial velocities in our analysis. The constraint provided by the requirement that astrometry and radial velocities describe the same physical system improves the accuracy of our result.

\section{The Wolf 1062 Astrometric Reference Frame}  \label{AstRefs}
\begin{figure}[here]
\psfig{figure=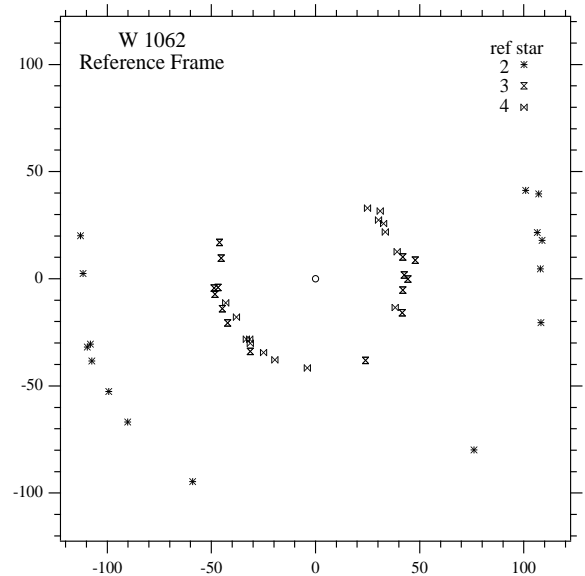,height=3.0in,width=3.0in}
\caption{ Wolf 1062 and reference
frame observations in FGS 3 pickle coordinates. The symbol shape
identifies each star listed in Table 3.} 
\label{fig-1}
\end{figure}

Figure \ref{fig-1} shows the distribution in FGS 3 pickle coordinates of the 15 sets of three reference star measurements for the Wolf 1062 reference frame. The circular pattern is impressed by the requirement that {\it HST} roll to keep its solar panels fully illuminated throughout the year. At each epoch we measured reference stars 2 and 4 once and reference star 3 twice for a total of 60 reference star observations.

\subsection{The Astrometric Model}
From these data we determine the scale, rotation, and offset ``plate
constants" relative to an arbitrarily adopted constraint epoch (the so-called ``master plate") for
each observation set. The Wolf 1062 reference frame contains only three stars. Hence, we constrain the scales along x and y to equality and the two axes to orthogonality. The consequences of this choice are minimal. For example, imposing these constraints on the Barnard's Star astrometry discussed in
Benedict et al (1999) results in an unchanged parallax and increases the error by 0.1 mas, compared to a full 6 parameter model.

The astrometric residuals for an initial solution constraining the parallax and proper motions of the three reference stars
to zero values showed that one reference star had non-zero proper motion. We therefore include the effects of reference star \#4 (ref-4) parallax ($\pi$) and proper motion ($\mu$). 
Our reference frame model becomes in terms of standard coordinates
\beq
\xi = ax + by + c -P_x*\pi - \mu_x*t 
\eeq 
\beq 
\eta = -bx +ay +f-P_y*\pi - \mu_y*t 
\eeq 

\noindent where a, b, c and f are the plate constants obtained from the reference frame, and t is time. We constrain $ \mu = 0$ and   
$ \pi = 0$ for reference stars ref-2 and ref-3. 
The master plate orientation to the sky is obtained from ground-based astrometry 
(USNO-A2.0 catalog, \cite{Mon98}) with uncertainties in the field orientation $\pm 0\fdg3$. We obtain the parallax factors, $P_x$ and $P_y$ from a JPL Earth orbit predictor (\cite{Sta90}), upgraded to version DE405. From USNO-A2.0 photometry (calibrated as discussed in \cite{Ben00a} and presented in Table 2) we see that the colors of the reference stars and the target are all fairly red. Hence, we apply no corrections for lateral color (Benedict et al 1999).

\subsection{Assessing Reference Frame Residuals}
\begin{figure}[here]
\psfig{figure=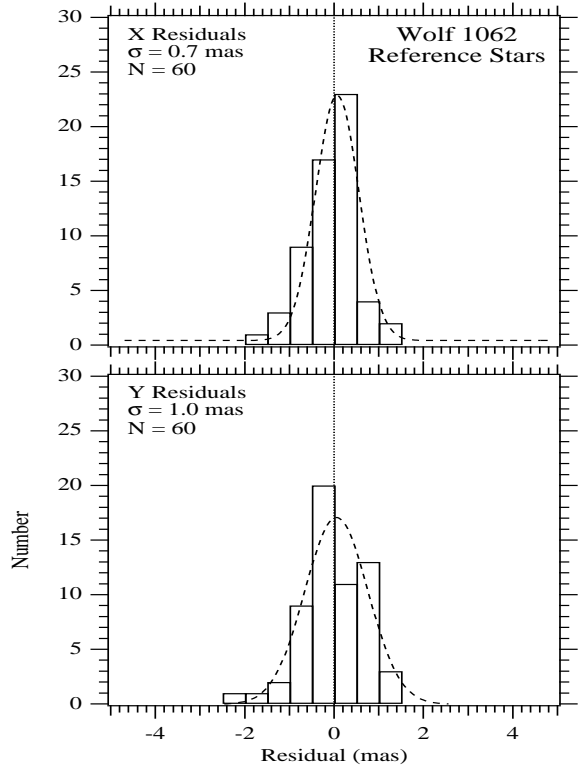,height=4.0in,width=3.0in}
\caption{Histograms of x and y residuals obtained from modeling the Wolf 1062 reference frame with equations 1 and 2. Distributions are fit
with gaussians.} \label{fig-2}
\end{figure}

The Optical Field Angle Distortion calibration (\cite{McA97}) reduces as-built {\it HST} telescope and FGS 3 distortions with magnitude $\sim1\arcsec$ to below 2 mas  over much of the FGS 3 field of regard. From histograms of the astrometric residuals (Figure~\ref{fig-2}) we conclude that we have obtained correction at the $\sim 1$ mas level in the
region available at all {\it HST} rolls (an inscribed circle centered on the pickle-shaped FGS field of regard). The resulting reference frame 'catalog' in $\xi$ and $\eta$ standard coordinates (Table \ref{tbl-2}) was determined
with	$<\sigma_\xi>= 0.5$	 and	$<\sigma_\eta> = 0.6$ mas.

To determine if there might be unmodeled - but possibly correctable -  systematic effects at the 1 mas level, we plotted the Wolf 1062 reference frame X and Y residuals against a number of spacecraft, instrumental, and astronomical parameters. These included X, Y position within the pickle; radial distance from the pickle center; reference star V magnitude and B-V color; and epoch of observation.  We saw no obvious trends, other than an expected increase in positional uncertainty with reference star magnitude. 

\section{Radial Velocities}
Our radial velocity data have two sources. We obtain three early epochs of absolute radial velocities from \cite{Mar89}.  To this we add radial velocity measurements (Table 5) at 12 epochs, obtained with the McDonald 2.1m telescope and Sandiford cassegrain echelle spectrograph (\cite{McC93}). The McDonald data were reduced using the standard IRAF (\cite{Tod93}) {\tt echelle} package tools, including {\tt fxcorr}. Treating Wolf 1062 as a double lined spectroscopic binary, we obtained velocities for both components at all but one orbital phase.

\section{ Wolf 1062 - Parallax and Orbits from Astrometry and Radial Velocities}

\subsection{Correction to Absolute Parallax}
Before we can determine the absolute masses of the A and B components, we need the distance, which we obtain via a parallax relative to our reference stars. Every small-field astrometric technique requires a correction from relative to absolute parallax because the reference frame stars have an intrinsic parallax. 
We adopt the corrections discussed and presented in the Yale Parallax Catalog (\cite{WvA95}, Section 3.2, Fig. 2, hereafter YPC95). We enter YPC95, Fig. 2, with the Wolf 1062 galactic latitude, $b = -3\fdg3$ and average magnitude for the reference frame, $<V_{ref}>=13.7$, and obtain a correction to absolute of 1.2 $\pm$ 0.2 mas. The error for the correction is determined by comparing the 
spectrophotometric parallaxes of the reference frames of the low galactic latitude science targets
discussed in \cite{Har99}, \cite{mca99}, and \cite{Ben99} with the YPC95 corrections.

\subsection{POS Mode Photocenter Corrections}
For binary stars the
FGS transforms the images of the two components into two overlapping fringes. For perfect fringes
and component separations greater than the resolution
of {\it HST} at $\lambda = 580$ nm, about 40 mas, the presence of a
companion should have little effect on the component A position obtained from the fringe
zero-crossing. However, FGS 3 does not produce perfect fringes (\cite{Fra98}). Given the uncorrected HST point spread
function (PSF) for the FGSs (they are not in the COSTAR path), we must apply
 photocenter-type corrections (e.g., \cite{vdK67}). The measurement of the position of the brighter component of a blended image will be biased towards the fainter. Because the detector has two orthogonal axes, the separation along 
each interferometer axis, not the total separation
on the sky, determines the photocenter correction. For any arbitrary {\it HST} orientation components A and B can have 
separations along the FGS axes from 0.0 mas to the actual
full separation. For example, observation set 17 in Table 4 has a total separation, $\rho_A=176.2$ mas. Resolved on each interferometer axis, $\Delta x = 52.3$ mas and $\Delta y = 168.2$ mas.

\begin{figure}[here]
\psfig{figure=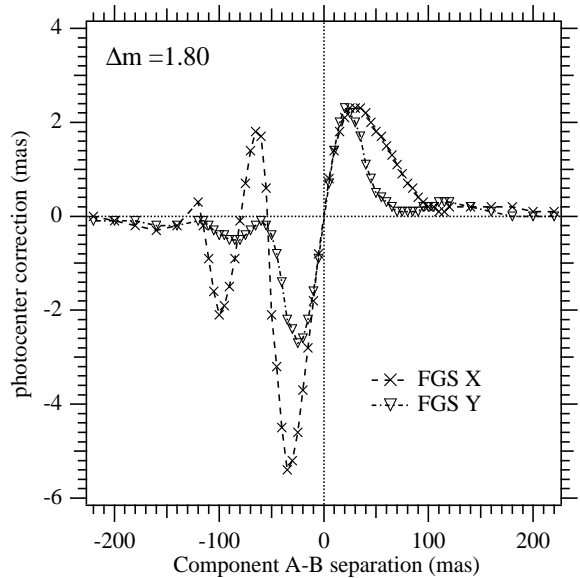,height=3.0in,width=3.0in}
\caption{ Fringe tracking (POS) photocenter corrections in mas for various component A - B separations along the FGS X and Y axes. Y corrections are, as expected, more symmetric and smaller.} 
\label{fig-3}
\end{figure}

Simulations were done to characterize the effect the secondary has
 on the position derived for the primary for separations
less than 200 mas.  First, we obtained
a position (the central zero-crossing point of the fringe) derived from a fringe scan of a single star of appropriate color.
Then, a companion with $\Delta$V = 1.80 (the magnitude difference between the components in this system) was placed at many separations along each axis and a combined fringe obtained. As can be seen in fig. 1 of Franz et al (1998) the interferometer response functions
are not symmetric about the zero crossing, particularly for the X fringe. Hence the simulation required
that the companion be placed on either side of the primary.
Fitting each synthetic binary TRANS scan as a single star will quantify the effect of the companion on the measured position of the primary.

The results of this simulation are shown in Figure~\ref{fig-3}. Classically, the correction always increases the separation of the two components. In general we
see this for both axes. However, such is the perversity of the X axis fringe that at times the correction works in the opposite sense. Photocenter 
corrections are negligible for separations greater than 120 mas for either axis. As expected from the differing X and Y fringe morphologies, the corrections are strongly asymmetric along the X axis, peaking at $\sim-5$~mas at a separation of -35 mas. The measures with corrections applied are listed in Table 3.

\subsection{Simultaneous Estimation of Parallax, Proper Motion, Mass Ratio, and Orbit Parameters for Wolf 1062}

We apply the transformations (equations 1 and 2) to the Wolf 1062 POS (corrected for photocenter effects) and TRANS measurements. We fold in the radial velocity measurements, solving for relative parallax, proper motion, and orbital motion of both components simultaneously. 
The model now becomes

\beq
\xi = aX + bY + c - P_x*\pi - \mu_x*t - ORBIT_x
\eeq 
\beq 
\eta = -bX +aY +f - P_y*\pi - \mu_y*t - ORBIT_y
\eeq 

\noindent ORBIT is a function of the traditional astrometric and radial velocity orbital elements, listed in Table 7.

For the simultaneous solution we constrain equality for the eccentricities (e) and longitudes of periastron ($\omega$) of the two component orbits. We also constrain the position angle of the line of nodes ($\Omega$) for the component A and B orbits to differ by 180\arcdeg. The period (P), the epoch of passage through periastron in years (T), the
eccentricity (e) and the angle in the plane of the true orbit between
the line of nodes and the major axis ($\omega$), are constrained to be equal
in the radial velocity and two modes of astrometry.  Only radial velocity
provides information with which to determine the half-amplitudes ($K_1$, $K_2$) and $\gamma$, the systemic velocity.  Combining
radial velocity observations from different sources is simple with GaussFit
(\cite{Jef87}),
which has the ability to simultaneously solve for two separate $\gamma$'s along with
the other orbital parameters. We easily mix absolute velocities from \cite{Mar89} with the relative velocities from our own program.

The relationship between the two astrometric modes (POS and TRANS)
and the radial velocity is
enforced by the constraint (\cite{Pou00})
\beq
\displaystyle{{\alpha_A~sin~i \over \pi_{abs}} = {P K_1 sqrt(1 -e^2)\over2\pi\times4.7405}} 
\eeq
\noindent where quantities derived only from astrometry (parallax, $\pi_{abs}$, primary perturbation orbit size, $\alpha_A$, and inclination, $i$) are on the left, and quantities derivable from both (the period, $P$ and eccentricity, $e$), or radial velocities only (the radial velocity amplitude for the primary, $K_1$), are on the right.

\begin{figure}[here]
\psfig{figure=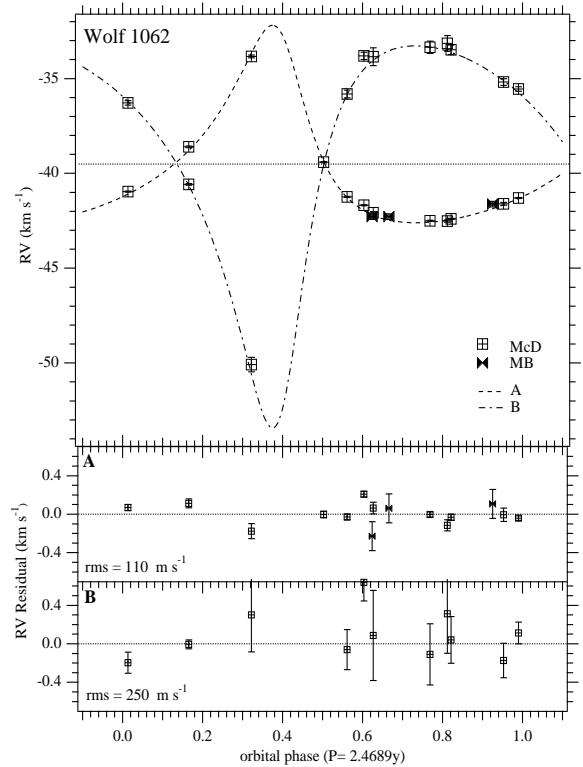,height=4.0in,width=3.0in}
\caption{ Top panel: radial velocity measurements from \cite{Mar89} (MB) and the present study (McD), phased to the orbital period determined from a combined solution including astrometry and radial velocity. The dashed lines are velocities
predicted from the orbital parameters derived in the combined solution. Middle and bottom panels: radial velocity residuals from the combined solution. The error bars on the residuals are the original measurement errors (1$\sigma$).} \label{fig-4}
\end{figure}

\begin{figure}[here]
\psfig{figure=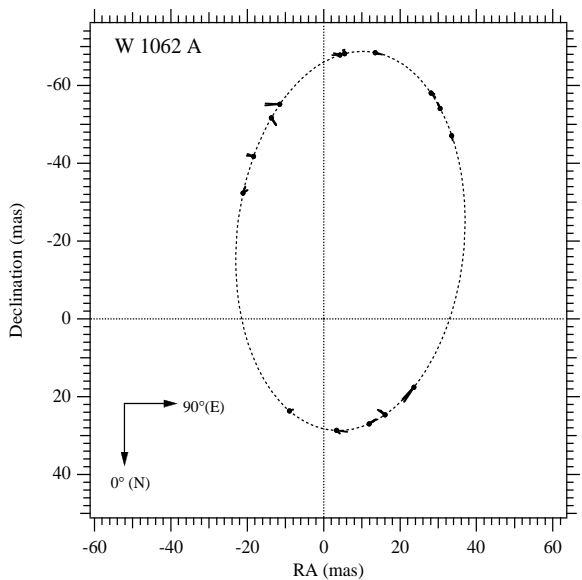,height=3.0in,width=3.0in}
\caption{Perturbation orbit for Wolf 1062 A. Elements are found in Table~\ref{tbl-6}. Dots are predicted
positions at each epoch of observation. Residual vectors are plotted
for each observation, two at each epoch; (see
Table~\ref{tbl-2}).} 
\label{fig-5}
\end{figure}

\begin{figure}[here]
\psfig{figure=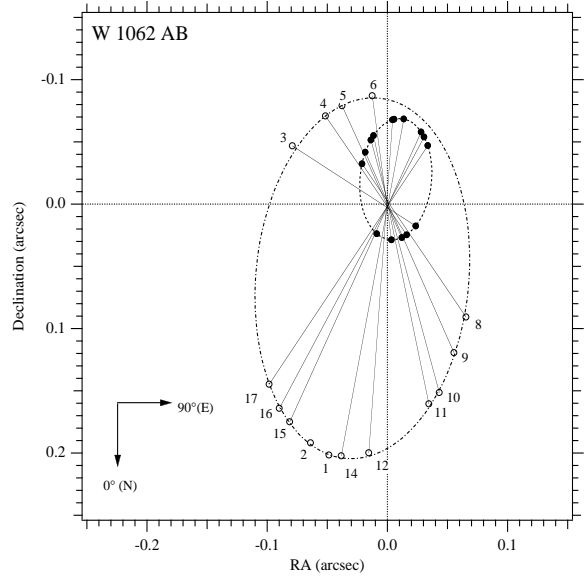,height=3.0in,width=3.0in}
\caption{ Wolf 1062 A (dots, POS measurements) and
component B (open circle, TRANS measures). Component B
positions are labeled with their corresponding observation set numbers
(Table~\ref{tbl-3}). All observations, POS and TRANS and A and B component radial velocities, were used to derive the Table 7 orbital elements. Points 1 and 2 had TRANS only. Points 7 and 13 were resolved only along one axis, hence not plotted.} 
\label{fig-6}
\end{figure}

The results of this simultaneous solution are as follows. The absolute parallax and the proper
motion are presented in Table~\ref{tbl-6} (errors are $1\sigma$). The final formal parallax uncertainty includes the estimated error in the correction to absolute, 0.2 mas,  combined in quadrature with the relative parallax error. Our precision has improved our knowledge of the parallax by a factor of six, comparing with HIPPARCOS and the Yale Parallax Catalog.
Table ~\ref{tbl-7} contains the relative orbital parameters with formal ($1 \sigma$) uncertainties. For comparison we include our previous relative orbit from \cite{Fra99}. Residuals for the POS, TRANS and radial velocity observations are given in Tables ~\ref{tbl-3}, ~\ref{tbl-4}, and ~\ref{tbl-5} respectively. Figure~\ref{fig-4} contains all radial velocity measures, radial velocity residuals, and the predicted velocity curve from the simultaneous solutions. 
Figure~\ref{fig-5} presents the astrometric residuals and the derived perturbation orbit for component A. Figure~\ref{fig-6} provides component A and component B orbits. The vectors connecting the components all pass through the center of mass with an
average positional scatter of less than 2 mas. 

\begin{figure}[here]
\psfig{figure=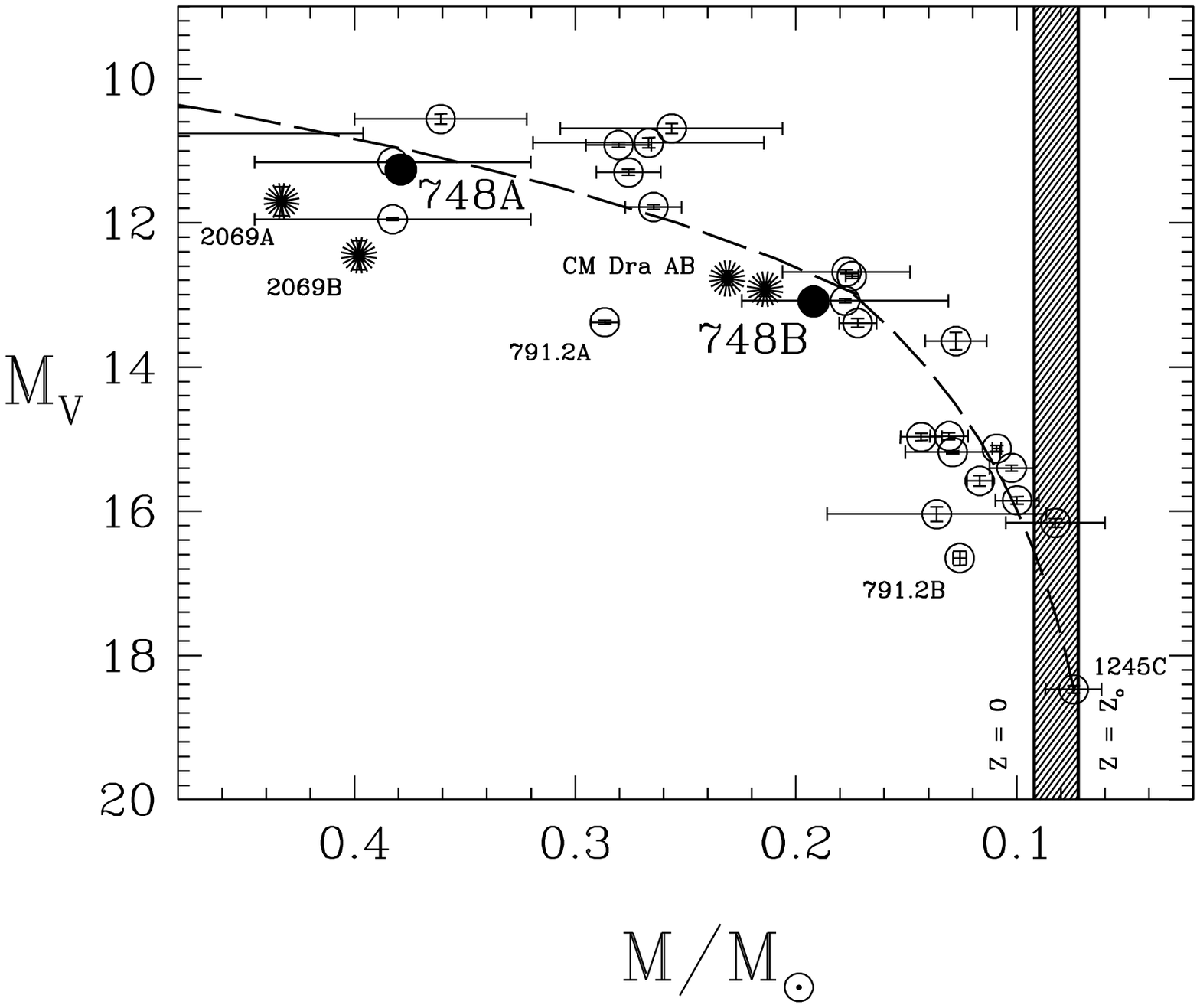,height=8.0in,width=6.0in}
\caption{ The components of Wolf 1062 are shown on the mass-luminosity diagram as solid
points.  The point size is
representative of the Wolf 1062 mass errors.  A few other stars of
interest are labeled, including the four components of the GJ 2069 (\cite{Del99}) and
CM Dra (\cite{Met96}) eclipsing binary systems (starred points) and the recent
determinations for Gl 791.2 AB by this group (Benedict et al. 2000b).  Also labeled is GJ
1245C, still the lowest mass object for which an accurate dynamical
mass has been determined.    
The dashed curve is the empirical mass-luminosity
relation from Henry and McCarthy (1993) down to 0.18 M$_{\odot}$ and
from Henry et al (1999) at lower masses.  The shaded region with
borders at 0.092 and 0.072 M$_{\odot}$ marks the main-sequence minimum
mass range for objects with zero to solar metallicity.} 
\label{fig-7}
\end{figure}

\section{Component Masses}
Our orbit solutions (Table~\ref{tbl-7}) and associated absolute parallax
(Table~\ref{tbl-6}) provide an orbit semi-major axis, $a$ in AU, from which we can determine the system mass through Kepler's Law.  
Given P and $a$, we solve the expression
			\beq
a^3 / P^2 =({\cal M}_A+ {\cal M}_B)= {\cal M}_{tot}
\eeq
We find ${\cal M}_{tot} = 0.571 \pm 0.008{\cal M}_{\sun}$.
At each instant
in the orbits of the two components around the common center of mass,
\beq
{\cal M}_A / {\cal M}_B = \alpha_B / \alpha_A
\eeq
\noindent a relationship that contains only one observable, $\alpha_A$, the perturbation orbit size. We, instead,  calculate the mass fraction
\beq
f = {\cal M}_B/({\cal M}_A+{\cal M}_B) = \alpha_A / (\alpha_A + \alpha_B) = \alpha_A /a,
\eeq
where $ \alpha_B = $a$-\alpha_A.$ 
This parameter ratios the two quantities directly obtained from the observations; the perturbation orbit size ($\alpha_A$ from POS mode) and the relative orbit size ($a$ from TRANS mode) both shown in Figure~\ref{fig-6} and listed in Table~\ref{tbl-7}. From these we derive a mass fraction of 0.3358 $\pm$ 0.0021. 
Equations 6, 7, and 8 yield ${\cal M}_A = 0.379 \pm 0.005{\cal M}_{\sun}$ and  ${\cal M}_B= 0.192 \pm 0.003 {\cal M}_{\sun}$, indicating that component B is of very low mass and is a valuable member of the 20-20-20 club with a mass error of only 1.5\%.

\section{Wolf 1062 A and B on the MLR}

With the component masses determined, we require the component
absolute magnitudes to place these stars on the MLR.  We use the photometry from \cite{Leg92},
$V_{tot} = 11.12 \pm 0.03$, and a $\Delta V$ = 1.83 $\pm$ 0.03 from Henry et al (1999) to derive component magnitudes V$_A$= 11.30$\pm$ 0.03 and V$_B$= 13.13$\pm$ 0.04.
With our parallax, $\pi_{abs} = 98.0 \pm 0.4$ mas we find
M$_{VA} = 11.26 \pm 0.03$ and M$_{VB} = 13.09\pm 0.04$.  For this very nearby system we
have assumed no absorption ($A_V = 0$). Neither does a parallax of this precision require correction for Lutz-Kelker bias (\cite{Lut73}) in the derived absolute magnitudes. We collect all derived mass
and absolute magnitude values in Table~\ref{tbl-8}.

Components A and B lie on the MLR as shown in Figure~\ref{fig-7}.
The lower main sequence relation is from Henry et al. (1999). The higher mass section (${\cal M} > 0.2{\cal M}_{\sun}$) is from Henry \& McCarthy (1993).

\section{Conclusions}

1. We obtained POS observations of the low-mass binary Wolf 1062 
over a 1.8 year time span and TRANS obtained over a 2.9 year span with {\it HST} FGS 3. These yield an absolute parallax $\pi_{abs}= $98.0 $\pm$ 0.4 mas with an external error better than 1\%.

2. Fringe tracking (POS) observations of the primary, component A, provide a
well-determined perturbation orbit, once photocenter effects are corrected. 

3. Fringe scans (TRANS) combined with POS observations provide a mass fraction =
0.3358 $\pm$ 0.0021 relative to an astrometric reference frame. TRANS also
provides a $\Delta V$ = 1.83 $\pm$ 0.03 (Henry et al 1999).

4. We obtain masses precise to 1.5\%: ${\cal M}_{tot} = 0.571 \pm
0.008{\cal M}_{\sun}$, ${\cal M}_A = 0.379 \pm 0.005{\cal M}_{\sun}$
and ${\cal M}_B= 0.192 \pm 0.003 {\cal M}_{\sun}$.

5. The system magnitude, the component magnitude difference, $\Delta V$, and our parallax provide component absolute magnitudes, M$_{VA} = 11.26 \pm 0.03$ and M$_{VB} = 13.09\pm 0.04$.

6. These masses and absolute magnitudes assist in
defining the lower main sequence MLR.  The primary provides one of the few  high quality masses between 0.2${\cal M}_{\sun}$ and 0.5${\cal M}_{\sun}$. The secondary lies below 0.2 M$_{\odot}$, the crucial region where age
begins to play a significant role in the flux of stars.

\acknowledgments

Support for this work was provided by NASA through grants GTO NAG5-
1603,
GO-06036.01-94A and GO-07491.01-97A from the Space Telescope 
Science Institute, which is operated
by the Association of Universities for Research in Astronomy, Inc., under
NASA contract NAS5-26555. We thank Linda Abramowicz-Reed, now
at Raytheon-Danbury, for her unflagging and expert instrumental support over
the last 15 years. Finally, we thank an anonymous referee who made several suggestions that resulted in  an improved paper.
\clearpage


\clearpage
\begin{center}
\begin{deluxetable}{lll}
\tablecaption{ Wolf 1062 = Gl 748 = HIP 94349 = LHS 472 = G 22-18
 \label{tbl-1}}
\tablewidth{0in}
\tablehead{\colhead{Parameter} &  \colhead{Value}&
\colhead{Reference}}
\startdata
V & 11.12$\pm$0.04 & \cite{Leg92} \\
\bv & 1.51 $\pm$ 0.05 & \cite{Leg92} \\
Sp.T. & M3.5V &    \cite{Hen99}\\
\enddata
\end{deluxetable}
\end{center}

\begin{center}
\begin{deluxetable}{llllllll}
\tablewidth{0in}
\tablecaption{Wolf 1062 Reference Frame  \label{tbl-2}}
\tablehead{  \colhead{ID}&
\colhead{V}&   \colhead{\bv}&  \colhead{$\xi$}&  
\colhead{$\eta$}&   \colhead{$\mu_X$} & \colhead{$\mu_Y$} &\colhead{$\pi$}\\
& & & (arcsec)&  (arcsec)& (arcsec y$^{-1}$)&(arcsec y$^{-1}$)&(arcsec)}
\startdata
\nl
ref-2*&12.8&0.7&0.0000$\pm$0.0006&0.0000$\pm$0.0007&0&0&0\nl
ref-3&14.1&1.4&-59.2177$\pm$0.0004&-30.8548$\pm$0.0006&0&0&0\nl
ref-4&14.2&0.7&-76.0673$\pm$0.0005&-7.8994$\pm$0.0007&-0.0131$\pm$0.0005&0.0034$\pm$0.0006&0.0002$\pm$0.0004\nl
\enddata
\tablenotetext{*} {RA, Dec = 288\fdg087967, 2\fdg898281 (J2000)}
\end{deluxetable}
\end{center}

\begin{center}
\begin{deluxetable}{lcccccc}
\tablecaption{POS Observations of Wolf 1062A and Perturbation Orbit Residuals \label{tbl-3}}
\tablewidth{0in}
\tablehead{\colhead{Obs. Set} &  \colhead{ MJD }&  \colhead{ $\rho_A$ (mas) }&  \colhead{ $\theta$ (\arcdeg) }&  \colhead{ $\Delta\rho_A$ (mas)}&  \colhead{$\Delta\theta$ (\arcdeg) }&  \colhead{ $\rho_A\Delta\theta$ (mas) }}
\scriptsize
\startdata
3&49926.481&32.6&53.3&-0.5&7.1&4.0\\
3&49926.499&32.9&53.3&0.1&6.0&3.4\\
4&49956.507&31.7&33.1&2.0&0.5&0.3\\
4&49956.525&31.2&33.1&1.6&0.0&0.0\\
5&49970.380&29.6&23.8&0.2&-3.5&-1.8\\
5&49970.397&29.7&23.8&0.2&-4.4&-2.3\\
6&49995.646&28.9&6.7&-0.8&-5.5&-2.8\\
6&49995.664&28.4&6.7&-0.9&-1.8&-0.9\\
7&50031.034&25.8&339.2&0.7&-0.7&-0.3\\
7&50031.052&25.0&339.2&0.1&0.7&0.3\\
8&50164.733&38.0&213.1&0.0&-2.0&-1.3\\
8&50164.751&39.0&213.1&-1.0&-1.8&-1.2\\
9&50193.22&45.7&203.7&-0.9&1.8&1.4\\
9&50193.238&46.1&203.7&-1.2&1.4&1.2\\
10&50230.679&52.2&194.9&2.1&-2.3&-2.1\\
10&50230.697&52.2&194.9&2.2&-2.3&-2.1\\
11&50246.763&53.4&191.8&-1.1&0.7&0.6\\
11&50246.780&53.1&191.8&-0.6&0.3&0.3\\
12&50351.175&66.6&176.4&0.2&-1.7&-2.0\\
12&50351.196&66.0&176.4&-0.4&-1.4&-1.6\\
13&50359.081&67.4&175.4&-0.9&-0.4&-0.5\\
13&50359.102&67.3&175.4&-1.0&-0.1&-0.1\\
14&50413.362&67.6&168.8&0.1&1.3&1.5\\
14&50413.384&68.1&168.8&0.2&0.7&0.9\\
15&50529.867&63.2&154.2&-0.2&-1.0&-1.1\\
15&50529.888&62.5&154.2&0.5&-1.6&-1.7\\
16&50554.807&61.0&150.6&-0.8&0.7&0.7\\
16&50554.828&61.0&150.6&-0.8&0.6&0.7\\
17&50592.184&58.1&144.5&0.8&-0.4&-0.4\\
17&50592.205&58.4&144.5&0.3&0.2&0.2\\
& N = 30& $<|res|>$& =  &0.8&1.8&1.3\\
\enddata
\end{deluxetable}
\end{center}

\clearpage
\begin{center}
\begin{deluxetable}{lcccccc}
\tablecaption{TRANS Observations of Wolf 1062B and Residuals \label{tbl-4}}
\tablewidth{0in}
\tablehead{\colhead{Obs. Set} &  \colhead{ MJD }&  \colhead{ $\rho_A$ (mas) }&  \colhead{ $\theta$ (\arcdeg) }&  \colhead{ $\Delta\rho_A$ (mas)}&  \colhead{$\Delta\theta$ (\arcdeg) }&  \colhead{ $\rho_A\Delta\theta$ (mas) }}
\scriptsize
\startdata
1&49534.723&207.5&346.5&-0.2&0.1&0.4\\
2&49578.078&201.5&341.6&0.8&-0.3&-1.0\\
3&49926.490&96.5&239.4&-4.6&-2.9&-5.0\\
4&49956.516&87.7&216&-0.1&0.0&0.0\\
5&49970.389&87.7&205.6&0.0&1.1&1.7\\
6&49995.655&89.7&188.3&-1.5&1.4&2.1\\
8&50164.742&111.9&35.7&-0.4&-0.9&-1.7\\
9&50193.229&130.5&24.9&0.9&0.1&0.2\\
10&50230.688&158.3&15.9&-1.0&-0.2&-0.5\\
11&50246.388&162.9&12.1&1.1&0.4&1.1\\
12&50351.186&199.5&355.6&0.9&1.4&4.7\\
14&50413.373&204.3&349.3&1.6&0.1&0.2\\
15&50529.878&193&335&-0.1&-0.3&-0.9\\
16&50554.818&187.9&331.2&-1.0&0.0&-0.1\\
17&50592.195&176.2&325.8&-1.4&-0.5&-1.5\\
 & N = 15& &$<|res|>=$&1.0&0.6&1.4\\ 
\enddata
\end{deluxetable}
\end{center}

\clearpage

\clearpage
\begin{center}
\begin{deluxetable}{lccrlcrlc}
\tablecaption{Radial Velocity Observations of Wolf 1062 and Residuals \label{tbl-5}}
\tablewidth{0in}
\tablehead{\colhead{Source\tablenotemark{a}} &  \colhead{MJD}&   \colhead{Phase}& \colhead{RV(A)\tablenotemark{b}}& &\colhead{Residual(A)} &\colhead{RV(B)\tablenotemark{b}}& &\colhead{Residual(B)}\\
&&&(km s$^{-1}$)& &(km s$^{-1}$)& (km s$^{-1}$)& &(km s$^{-1}$)}
\scriptsize
\startdata
MB&46627.5&0.699&-42.42&$\pm$0.30&-0.23& & &\\
MB&46665.5&0.741&-42.41&0.30&0.06&&&\\
MB&46899.5&0.000&-41.75&0.30&0.11&&&\\
McD&49960.65&0.395&-33.96&0.16&-0.18&-50.49&$\pm$0.77&0.30\\
McD&50214.42&0.676&-41.81&0.06&0.21&-34.21&0.39&0.64\\
McD&50364.96&0.843&-42.64&0.05&0.00&-33.75&0.64&-0.11\\
McD&50403.59&0.886&-42.64&0.11&-0.12&-33.55&0.82&0.31\\
McD&50530.01&0.026&-41.73&0.14&-0.01&-35.58&0.36&-0.17\\
McD&50584.56&0.087&-41.10&0.06&0.07&-36.69&0.22&-0.20\\
McD&50723.09&0.240&-38.74&0.09&0.11&-40.98&0.09&0.00\\
McD&51028.80&0.579&-39.54&0.07&0.00&&&\\
McD&51079.69&0.636&-41.37&0.06&-0.03&-36.21&0.42&-0.06\\
McD&51138.58&0.701&-42.22&0.12&0.06&-34.25&0.93&0.09\\
McD&51313.70&0.895&-42.53&0.06&-0.03&-33.88&0.49&0.04\\
McD&51466.12&0.064&-41.43&0.05&-0.04&-35.94&0.22&0.11\\
& N = 15& &&$<|res|>=$&0.08 & & &0.14\\ 
\enddata
\tablenotetext{*} {MB = Marcy \& Benitz 1989; McD = McDonald Observatory}
\tablenotetext{b} {The McD RV have been adjusted to the Marcy system (absolute velocities) using the two $\gamma$'s derived from the simultaneous solution.}
\end{deluxetable}
\end{center}

\begin{center}
\begin{deluxetable}{ll}
\tablecaption{Wolf 1062 Parallax and Proper Motion \label{tbl-6}}
\tablewidth{0in}
\tablehead{\colhead{Parameter} &  \colhead{ Value }}
\startdata
{\it HST} study duration  &1.8 y\nl
number of observation sets    &   15 \nl
ref. stars $ <V> $ &  $13.7 \pm 0.25$  \nl
ref. stars $ <B-V> $ &  $0.9 \pm 0.4$ \nl
\nl
{\it HST} Relative Parallax & 96.8 $\pm$ 0.3  mas\nl
correction to absolute & 1.2 $\pm$ 0.2   mas\nl
{\it HST} Absolute Parallax   & 98.0 $\pm$ 0.4   mas\nl
{\it HIPPARCOS} Absolute Parallax &98.6 $\pm$2.7 mas\nl
YPC95 Absolute Parallax & 99.8 $\pm$ 2.4  mas\nl
\nl
{\it HST} Proper Motion  &1856.5 $\pm$ 1.0 mas y$^{-1}$ \nl
 \indent in pos. angle & 106\fdg9 $\pm$0\fdg 1 \nl
HIPPARCOS Proper Motion  &1863.3 $\pm$ 3.4 mas y$^{-1}$ \nl
 \indent in pos angle & 106\fdg2 $\pm$0\fdg4 \nl
YPC95 Proper Motion  &1801.0  mas y$^{-1}$ \nl
 \indent in pos. angle & 105\fdg8  \nl

\enddata
\end{deluxetable}
\end{center}

\begin{center}
\begin{deluxetable}{llll}
\tablecaption{ Wolf 1062 Orbits \label{tbl-7}}
\tablewidth{0in}
\tablehead{\colhead{Parameter} &  \colhead{Component A}&  \colhead{Component B}& \colhead {Relative*}}
\startdata
$\alpha$(mas)&49.9 $\pm$ 0.3&98.6 $\pm$ 0.4&\nl
P(days)& 901.77 $\pm$ 0.47&-&900.9223 $\pm$ 3.0639\nl
P(years)& 2.4689 $\pm$ 0.0013&-&2.4664 $\pm$ 0.0075\nl
T$_0$ & 1995.8623 $\pm$ 0.0010&-&1995.8646 $\pm$ 0.0037\nl
e& 0.4519 $\pm$ 0.0013&-&0.4532 $\pm$ 0.0041\nl
i& 130\fdg8 $\pm$ 0\fdg3&-&131\fdg5 $\pm$ 0\fdg8\nl
$\Omega$&0\fdg7 $\pm$ 0\fdg1&180\fdg7 $\pm$ 0\fdg1&178\fdg8 $\pm$ 0\fdg6\nl
$\omega$&27\fdg3 $\pm$ 0\fdg1&-&26\fdg5 $\pm$ 1\fdg\nl
$K_1$&5.20 $\pm$ 0.06 \kms& & \nl
$K_2$&10.08 $\pm$ 0.19 \kms& & \nl
$\gamma$&39.5 $\pm$ 0.1 \kms& & \nl
\nl
$a$ (mas)& 148.5 $\pm$ 0.3& &147.95 $\pm$ 0.9\nl
$a$ (AU) & 1.515 $\pm$ 0.007 & & \nl
f& 0.3358 $\pm$ 0.0021& &
\enddata
\tablenotetext{*} {\cite{Fra99}}
\end{deluxetable}
\end{center}

\begin{center}
\begin{deluxetable}{cc}
\tablecaption{Wolf 1062: Component Masses and M$_V$ \label{tbl-8}}
\tablewidth{0in}
\tablehead{\colhead{Parameter} &  \colhead{Value}}
\startdata
${\cal M}_{tot} $&$ 0.571 \pm 0.008~{\cal M}_{\sun}$\\
${\cal M}_A $&$ 0.379 \pm 0.005~{\cal M}_{\sun}$\\
${\cal M}_B$&$ 0.192 \pm 0.003~{\cal M}_{\sun}$\\
$M_{VA} $&$ 11.26 \pm 0.03 $\\
$M_{VB} $&$13.09 \pm 0.04$\
\enddata
\end{deluxetable}
\end{center}

%
%

\end{document}